\documentclass[aps,prl,twocolumn,superscriptaddress,showpacs]{revtex4-1}
\usepackage{graphicx}

\RequirePackage{lineno}
\usepackage{hyperref}
\usepackage{textgreek}
\usepackage{booktabs}
\usepackage{color,soul}
\usepackage{xspace}
\newcommand{\nuGeN} {{{$\mathrm{\nu}$GeN }}}
\newcommand{\CEvNS} {{CE{$\mathrm{\nu}$NS }}}

\begin{document}

\title{First results of the \nuGeN experiment on coherent elastic neutrino-nucleus scattering}

\author{I.~Alekseev}
\affiliation{Dzhelepov Laboratory of Nuclear Problems, JINR, Joliot-Curie 6, 141980 Dubna, Russia}

\author{K.~Balej}
\affiliation{Czech Technical University in Prague, Institute of Experimental and Applied Physics, Prague, Czech Republic}
\affiliation{Charles University, Faculty of Mathematics and Physics, Prague, Czech Republic}

\author{ V.~Belov}
\affiliation{Dzhelepov Laboratory of Nuclear Problems, JINR, Joliot-Curie 6, 141980 Dubna, Russia}

\author{S.~Evseev}
\affiliation{Dzhelepov Laboratory of Nuclear Problems, JINR, Joliot-Curie 6, 141980 Dubna, Russia}

\author{D.~Filosofov}
\affiliation{Dzhelepov Laboratory of Nuclear Problems, JINR, Joliot-Curie 6, 141980 Dubna, Russia}

\author{M.~Fomina}
\affiliation{Dzhelepov Laboratory of Nuclear Problems, JINR, Joliot-Curie 6, 141980 Dubna, Russia}

\author{Z.~Hons}
\affiliation{Dzhelepov Laboratory of Nuclear Problems, JINR, Joliot-Curie 6, 141980 Dubna, Russia}
\affiliation{Czech Technical University in Prague, Institute of Experimental and Applied Physics, Prague, Czech Republic}

\author{D.~Karaivanov}
\affiliation{Dzhelepov Laboratory of Nuclear Problems, JINR, Joliot-Curie 6, 141980 Dubna, Russia}
\affiliation{ Institute for Nuclear Research and Nuclear Energy, 72 Tzarigradsko chaussee, Blvd., BG – 1784 Sofia, Bulgaria}

\author{S.~Kazartsev}
\affiliation{Dzhelepov Laboratory of Nuclear Problems, JINR, Joliot-Curie 6, 141980 Dubna, Russia}

\author{ J.~Khushvaktov}
\affiliation{Dzhelepov Laboratory of Nuclear Problems, JINR, Joliot-Curie 6, 141980 Dubna, Russia}

\author{A.~Kuznetsov}
\affiliation{Dzhelepov Laboratory of Nuclear Problems, JINR, Joliot-Curie 6, 141980 Dubna, Russia}

\author{A.~Lubashevskiy}
\email[]{lubashev@jinr.ru}
\affiliation{Dzhelepov Laboratory of Nuclear Problems, JINR, Joliot-Curie 6, 141980 Dubna, Russia}

\author{D.~Medvedev}
\affiliation{Dzhelepov Laboratory of Nuclear Problems, JINR, Joliot-Curie 6, 141980 Dubna, Russia}

\author{D.~Ponomarev}
\affiliation{Dzhelepov Laboratory of Nuclear Problems, JINR, Joliot-Curie 6, 141980 Dubna, Russia}

\author{A.~Rakhimov}
\affiliation{Dzhelepov Laboratory of Nuclear Problems, JINR, Joliot-Curie 6, 141980 Dubna, Russia}
\affiliation{Institute of Nuclear Physics of Uzbekistan Academy of Sciences (INP ASRUz), Tashkent, 100214, Uzbekistan}

\author{K.~Shakhov}
\affiliation{Dzhelepov Laboratory of Nuclear Problems, JINR, Joliot-Curie 6, 141980 Dubna, Russia}

\author{ E.~Shevchik}
\affiliation{Dzhelepov Laboratory of Nuclear Problems, JINR, Joliot-Curie 6, 141980 Dubna, Russia}

\author{M.~Shirchenko}
\affiliation{Dzhelepov Laboratory of Nuclear Problems, JINR, Joliot-Curie 6, 141980 Dubna, Russia}

\author{K.~Smolek}
\affiliation{Czech Technical University in Prague, Institute of Experimental and Applied Physics, Prague, Czech Republic}

\author{S.~Rozov}
\affiliation{Dzhelepov Laboratory of Nuclear Problems, JINR, Joliot-Curie 6, 141980 Dubna, Russia}

\author{I.~Rozova}
\affiliation{Dzhelepov Laboratory of Nuclear Problems, JINR, Joliot-Curie 6, 141980 Dubna, Russia}

\author{S.~Vasilyev}
\affiliation{Dzhelepov Laboratory of Nuclear Problems, JINR, Joliot-Curie 6, 141980 Dubna, Russia}

\author{E.~Yakushev}
\affiliation{Dzhelepov Laboratory of Nuclear Problems, JINR, Joliot-Curie 6, 141980 Dubna, Russia}

\author{I.~Zhitnikov}
\affiliation{Dzhelepov Laboratory of Nuclear Problems, JINR, Joliot-Curie 6, 141980 Dubna, Russia}

\collaboration{\nuGeN collaboration}

\date{\today}

\begin{abstract}
The \nuGeN experiment is aimed to investigate neutrino properties using antineutrinos from the reactor of the Kalinin Nuclear Power Plant. The experimental setup is located at about 11 meters from the center of the 3.1 GW$_{th}$ reactor core. Scattering of the antineutrinos from the reactor is detected with low energy threshold high purity germanium detector. Passive and active shieldings are used to suppress all kinds of backgrounds coming from surrounding materials and cosmic radiation. The description of the experimental setup together with the first results is presented. The data taken in regimes with reactor ON (94.50 days) and reactor OFF (47.09 days) have been compared. No significant difference between spectra of two data sets is observed, i.e. no positive signals for coherent elastic neutrino-nucleus scattering are detected. Under Standard Model assumptions about coherent neutrino scattering an upper limit on a quenching parameter k $<$ 0.26 (90 \% C.L.) in germanium has been set.
\end{abstract}


\maketitle

Neutrino is one of the most mysterious particles in modern physics. Precise knowledge of its properties is important for particle physics, astrophysics, and cosmology. To investigate fundamental neutrino parameters, it is required to have simultaneously a very strong source of neutrinos, advanced detection methods, and apply various methods for the suppression of background events.

The coherent elastic neutrino-nucleus scattering (CE{$\nu$NS) is a process predicted by the Standard Model~\cite{Fre74},~\cite{Dru84}. For a spin-zero mass of the nucleus $M$ the differential cross-section can be expressed as~\cite{And11}:
\begin{equation}
\Big(\frac{d\sigma}{dT}\Big) = \frac{G^{2}_{F}}{4 \pi} Q^2_W M \Big[1 - \frac{M T}{2E_{\nu}^2} \Big] F^2(Q^2),
\end{equation}
where $T$ is the nuclear recoil energy, $E_{\nu}$ is the neutrino energy, $Q$ is the transferred momentum, $F(Q^{2})$ is the nuclear form-factor, and $M$ is the mass of the target nucleus, $G_{F}$ is Fermi constant and $Q_W = N - (1-4 sin^2 \theta_W)$ is a weak charge. Due to the small momentum transfer, neutrino interacts simultaneously with all nucleons, and its cross-section is enhanced by several orders of magnitude in comparison with other neutrino interactions at the same energy. Since the predicted value of Weinberg angle at low energies is $sin^2 \theta_W = 0.23867\pm0.00016$~\cite{angleW}, the cross-section $\sigma$ is increased proportionally to the number of nuclear target neutrons squared $N^2$.

The COHERENT experiment reported an observation of \CEvNS using neutrino emissions produced by SNS accelerator~\cite{COHERENT},~\cite{COHERENT2}. At the neutrino energies of $\sim$50 MeV, the resulting signal includes partly the incoherent scattering~\cite{Bed18}. Therefore, the investigation of the low-energy anti-neutrinos from nuclear reactor can provide a precise information about the coherent elastic scattering. This is a key to the search for non-standard neutrino interactions, test of the Standard Model, tests in nuclear physics, and other investigations~\cite{Rev22}. The search for other effects, like the neutrino magnetic moment~\cite{Bed13} can be performed at the same time. 

The key factors needed for \CEvNS detection are a high neutrino flux, a low background level, a big mass of the target, and a low energy threshold. For example, for semiconductor germanium detector, it is necessary to detect signals below 0.4 keV, at the rates below a few counts per month. The discrimination from the noise of the low energy signals produced by nuclear recoils is one of the main experimental challenges. Furthermore, signals can be mimicked by some type of background, for example by neutron scattering. 

Extensive scientific efforts are focused on the search for \CEvNS in close vicinity of commercial power or experimental nuclear reactors. Different experimental techniques are used to detect and investigate CE{$\mathrm{\nu}$NS. There are several experiments, which are currently running or under construction: CONUS~\cite{Bon21}, Ricochet~\cite{Bea21}, NUCLEUS~\cite{Ang19}, CONNIE~\cite{Ale19} and many others. Currently, the sensitivities of these experiments are approaching to the possibility of \CEvNS detection.

The \nuGeN experimental setup is located at Kalinin Nuclear Power Plant (KNPP) in Udomlya, Russia, near the 3.1 GW$_{th}$ reactor unit \#3 of WWER-1000 type~\cite{Bel15}. The current distance from the detector to the center of the reactor core is 11.835 m, where the neutrino flux is 3.9$\cdot10^{13}$ $\mathrm{cm^{-2} s^{-1}}$ according to calculation method in~\cite{Bed07}. The closer position to the reactor core with a higher neutrino flux is planned to be explored as well. This neutrino intensity is several times higher in comparison to other groups worldwide, except the sideway site of the Dresden~II reactor~\cite{Col21}. Moreover, the available place at KNPP is located just under the reactor, which together with surrounding materials provides about 50 m w.e. shielding from cosmic rays~\footnote{The detailed investigation of the identical room at the similar reactor core unit \#4 are described in~\cite{Ale16}}.

Custom-designed high purity germanium detector made by Mirion Technologies (Canberra Lingolsheim)~\cite{Mir20} is used to detect CE{$\mathrm{\nu}$NS. It has been specially produced to achieve energy threshold as low as possible by taking into account low-radioactivity requirements. First measurements at KNPP were performed with the Ge detector with an active mass of 1.41 kg. The germanium crystal has a cylindrical shape with a diameter of 70 mm and height of 70 mm. The detector is installed inside the cryostat made of low background aluminum and copper. It is equipped with an electrically powered pulse tube cooler model Cryo-Pulse 5 Plus~\cite{Cry20}. The cooling temperature of the detector was optimized to the value of $-185^{\circ}$C.

The system of passive and active shielding has been built around the detector to suppress ambient background (see figure~\ref{fig:shielding}).
\begin{figure}
  \begin{center}
    \includegraphics[width=8.6 cm]{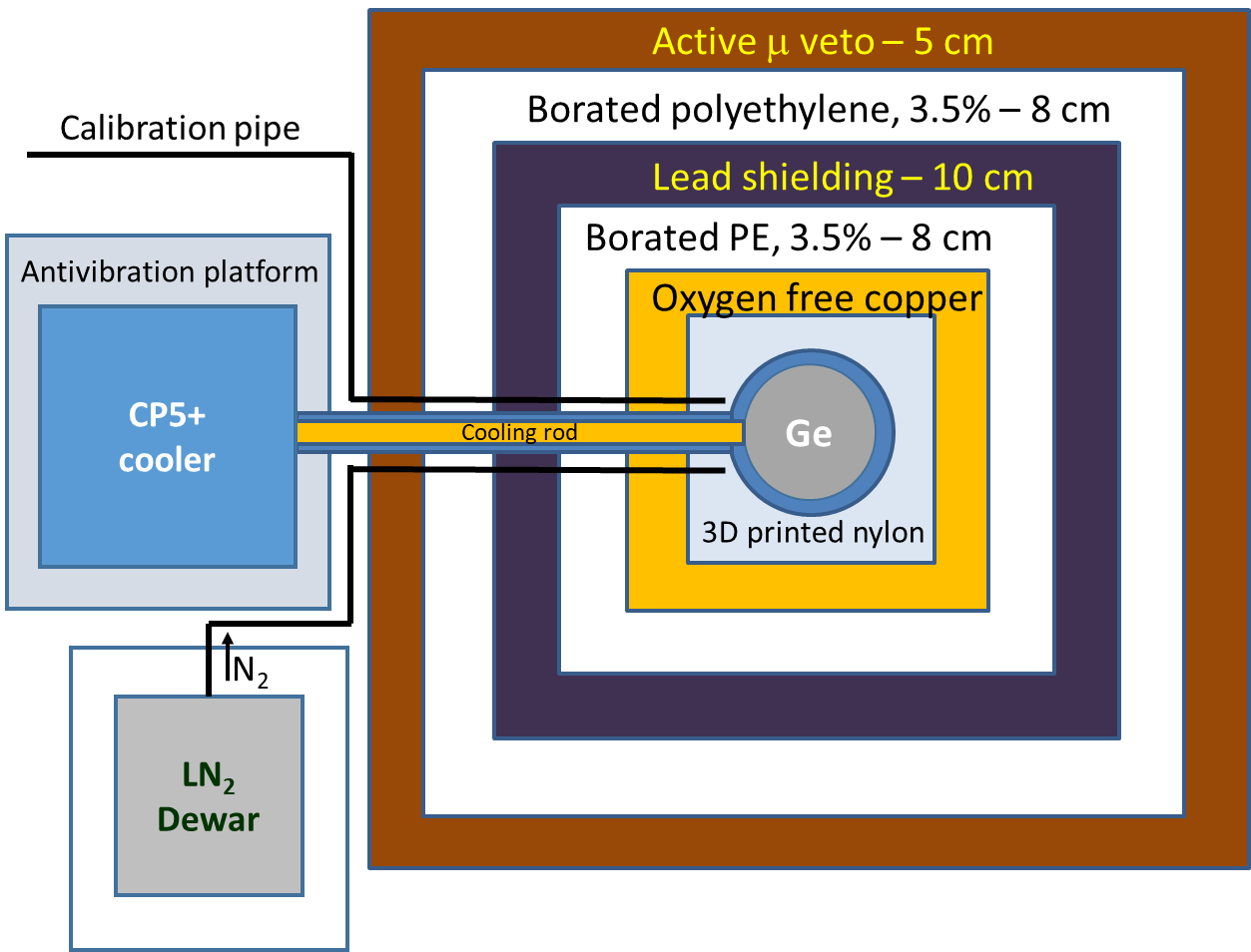}
    \caption{\label{fig:shielding} Scheme of the \nuGeN shielding. Top view.}
  \end{center}
\end{figure}
The most inner part of the shielding is made of 3D printed nylon in order to get rid of radon. The further layers are 10 cm of oxygen-free copper, 8 cm of borated (3.5\%) polyethylene, 10 cm layer of lead, another layer of 8 cm of borated polyethylene, and a 5 cm thick active muon veto made from plastic scintillator panels. Radon level inside the shielding is further decreased with the help of expulsion by nitrogen. The experimental site has various vibrations coming from the reactor equipment. Therefore, the detector is placed on an active anti-vibration platform TS-C30~\cite{Vib22}.

Ionization energy losses inside the HPGe detector result in a charge, collected on the electrodes. The charge is converted into voltage amplitude pulses by integrated cold and warm electronics. The electronic feedback resets the accumulated charge after a certain level. Even in near-zero background conditions, i.e. without signals, the leakage current through the detector causes the output to drift, requiring a reset. Thus, the reset frequency depends on the sum of the detector leakage current and the counting rate. For the \nuGeN detector, the reset rate is about 5 Hz. Figure~\ref{fig:acq} shows a diagram of the components involved in data acquisition.
\begin{figure}
  \begin{center}
    \includegraphics[width=8.6 cm]{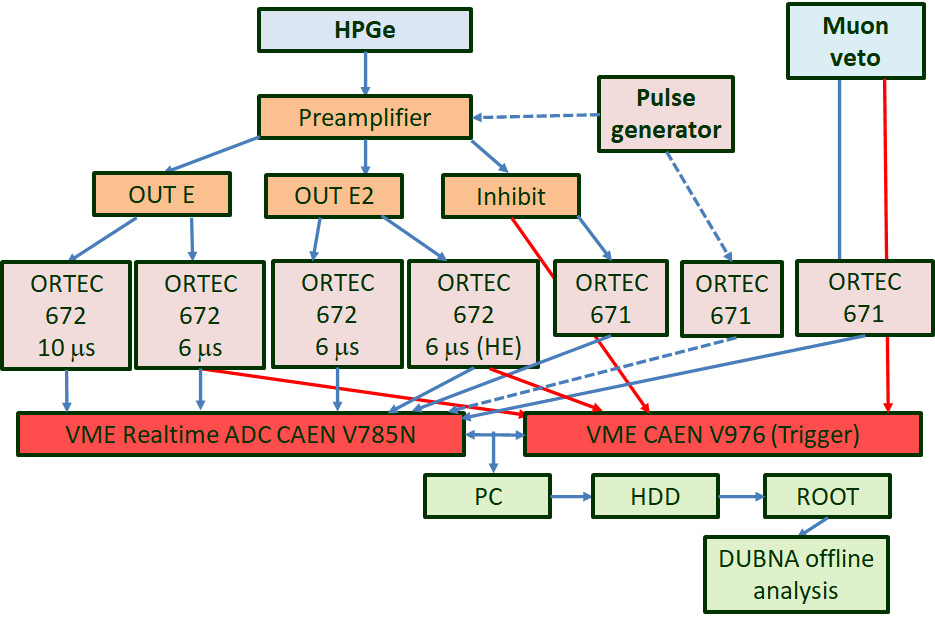}
    \caption{\label{fig:acq} Scheme of the acquisition system.}
  \end{center}
\end{figure}
The preamplification cascade is equipped with two similar amplitude outputs (OUT E and OUT E2). Additionally, during the reset, a logical Inhibit signal is generated. Information about each event corresponding to its energy together with timestamps is produced by the multichannel analog to digital converter CAEN VME Realtime ADC V785N. A positive signal suitable for use with the ADC is provided by spectroscopic shaping amplifiers. The primary purpose of the \nuGeN experiment is to perform spectroscopic measurements in the energy region where {CE{$\nu$NS signal can be detected, i.e. few hundred eV range (electron energy scale). During a preliminary set of measurements, it has been found that the lowest energy threshold and best energy resolution are achieved with 6 $\mu$s shaping time. Each of the E outputs of the preamplifier is connected to two ORTEC 672 spectroscopic amplifiers (four in total). The main energy spectrum is a result of averaged signals processed by amplifiers with the 6 $\mu$s shaping tuned for measurements below $\sim$13 keV. Comparison of the signals obtained with different amplifiers (same and different shaping times -- 6 and 10 $\mu$s) allows efficient noise discrimination~\cite{Mor92},~\cite{Ste02}. Additionally, a wide energy region up to 700 keV is measured with ORTEC 672 amplifier (labeled HE at figure~\ref{fig:acq}). Timestamps from the muon veto system and Inhibit signals are processed by the same ADC. CAEN VME V976 trigger unit issues an acquisition command on input conditions corresponding to: 1) low energy HPGe signal; 2) high energy HPGe signal; 3) inhibit logical signal; 4) muon veto. The acquisition software has been previously designed and commissioned for the DANSS experiment~\cite{Hons17},~\cite{Hons15}.

Due to KNPP safety restrictions, the experimental hall does not have Internet access, so the data is periodically copied for offline analysis. The setup has supplementary detectors used to control neutron and $\gamma$- backgrounds. The temperature and humidity in the room are continuously recorded.

The measurements have been carried out since 2020. Various tests, calibrations, and characterizations have been performed. A high-energy part of the spectrum is calibrated with the help of tungsten welding rods, which contains about 2\% of natural thorium. The energy calibration for the low energy part of the spectrum is determined with a help of the cosmogenic 10.37 keV line of $^{68,71}$Ge and pulse generators CAEN Mod.NTD6800D and ORTEC 419. The calibration at the low energy part of the spectrum is verified with help of the position of the 1.3 keV cosmogenic line of  $^{68,71}$Ge. The measurements demonstrated linearity of the energy scale (see figure~\ref{fig:calib}).
\begin{figure}
  \begin{center}
    \includegraphics[width=8.6 cm]{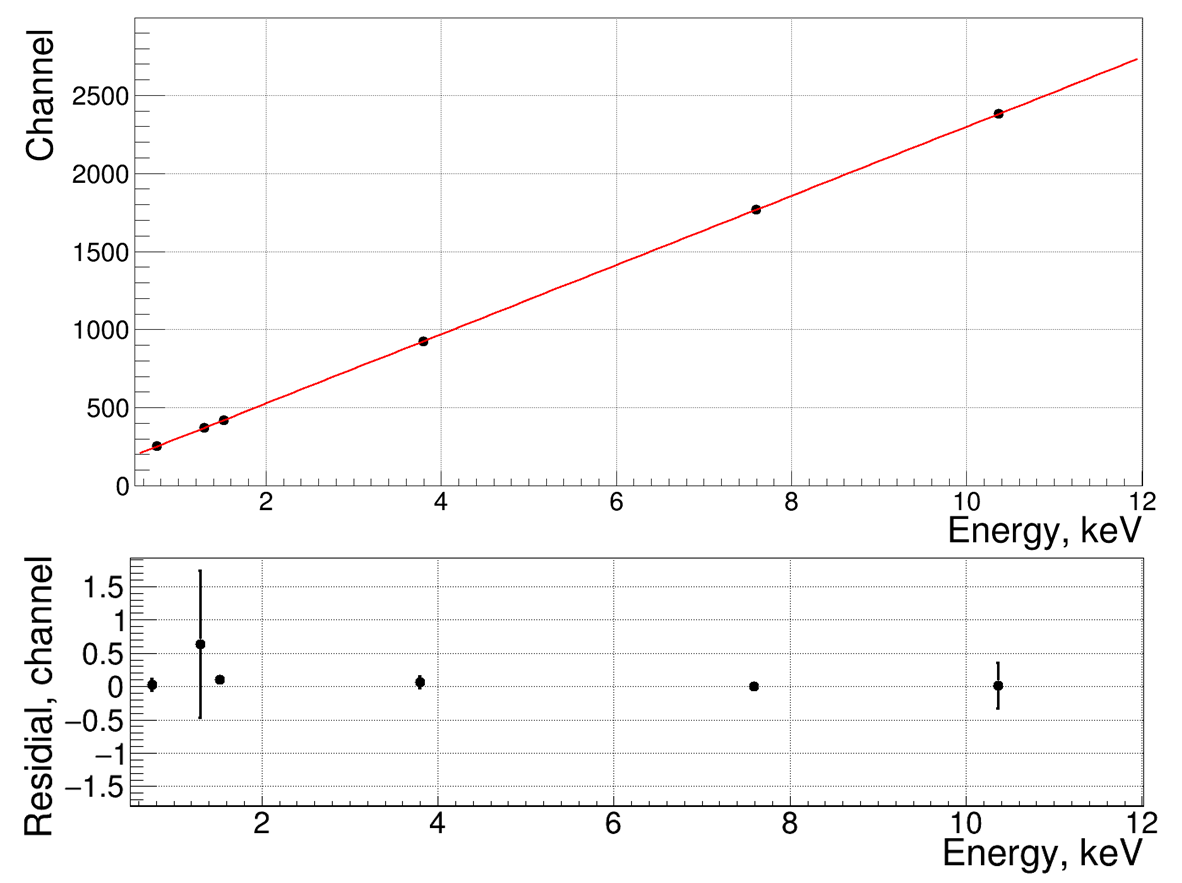}
    \caption{\label{fig:calib} Energy calibration of one of the channels performed with the pulse generator and cosmogenic line. Position of 1.298 keV line is used for verification of the calibration.}
\end{center}
\end{figure}
The energy resolution obtained with the pulse generator is 101.6$\pm$0.5 eV (FWHM).

The reset of the baseline produces small afterpulses, which can be interpreted as a physical signal. Therefore, the time period of 4.2 msec after each of the resets are not considered for the analysis. In addition, we look at the time intervals between two consecutive events to reduce the influence of microphonic noise. Comparison and averaging of the signals reconstructed with two different preamplifier outputs allows to suppress the noise appearing in the electronic tract. 

The efficiency of cuts application at low energy region in combination with a trigger efficiency has been tested by using the pulse generator. Depending on the energy of generated signals, it has been found that the detection efficiency is always higher than 80\% for signals above 0.25 keV. For 0.3 keV signals it is about 95\%, see figure~\ref{fig:eff}.
\begin{figure}
  \begin{center}
    \includegraphics[width=1\linewidth]{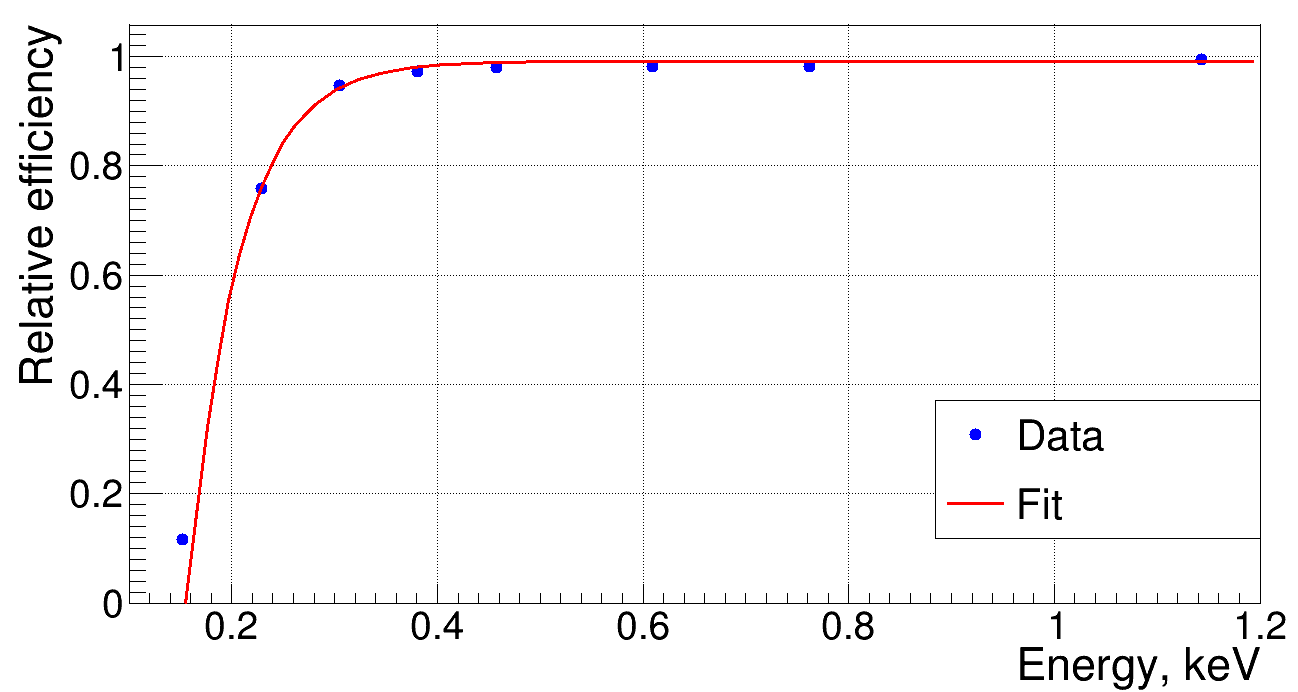}
    \caption{\label{fig:eff} Efficiency of signal detection measured with the pulse generator.}
\end{center}
\end{figure}
Since the region of analysis is above 0.3 keV it provides practically constant efficiency for all analyzed events. The energy spectrum before and after application of quality cuts and anti-coincidences with muon veto is shown in figure~\ref{fig:espectrum}.
\begin{figure}
  \begin{center}
    \includegraphics[width=8.6 cm]{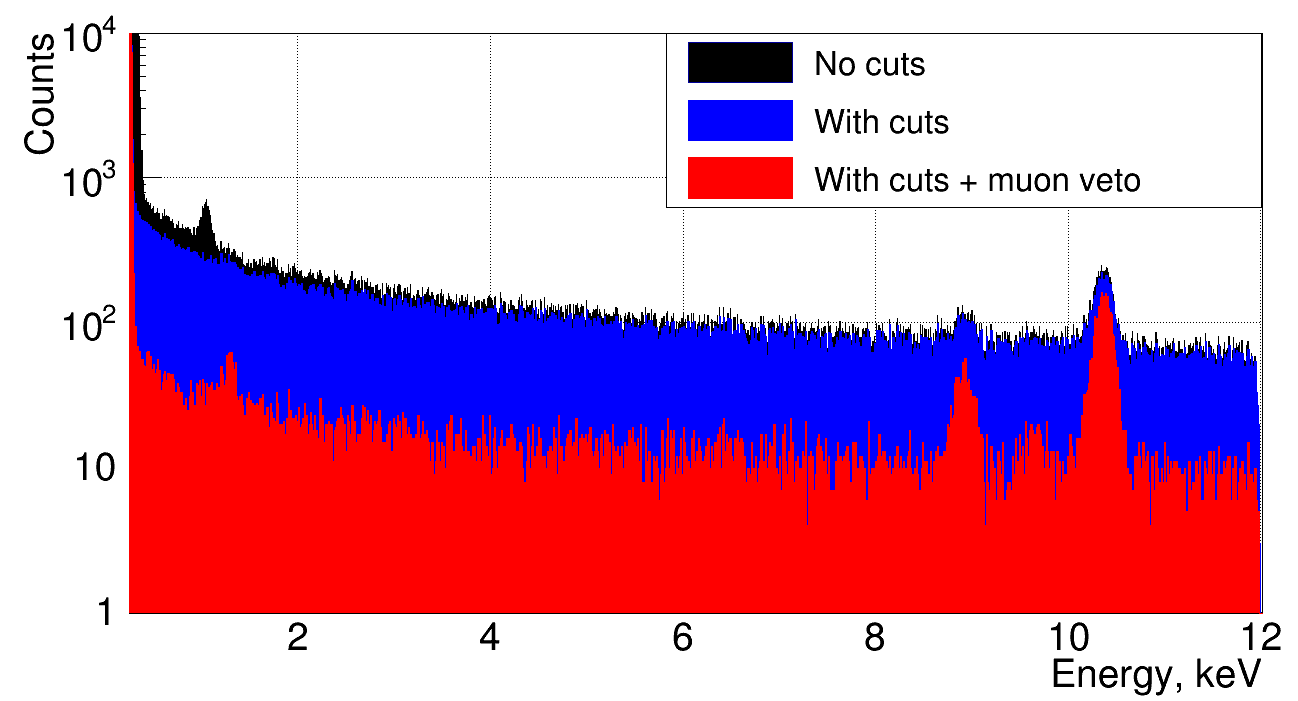}
    \caption{\label{fig:espectrum} Energy spectrum before (black) and after application of quality cuts (blue) and anti-coincidences with muon veto (red).}
\end{center}
\end{figure}
The change in the intensity of the 10.37 keV line allows to determine efficiency of cuts and muon veto. The obtained value of efficiency is 83.4$\pm$2.5\%.

Dedicated measurements for the search of the \CEvNS has been conducted since the end of 2020. The stability of data taking is an important factor that affects further data interpretation. The experimental hall is equipped with air conditioners to provide a stable temperature within $\pm$1$^{\circ}$C. The stability of the energy scale is verified by periodical Th calibration, the position of the cosmogenic 10.37 keV line, and calibrations with pulse generators. The accumulated days of the measurements and the thermal power of the reactor are shown in figure~\ref{fig:power}.
\begin{figure}
  \begin{center}
    \includegraphics[width=8.6 cm]{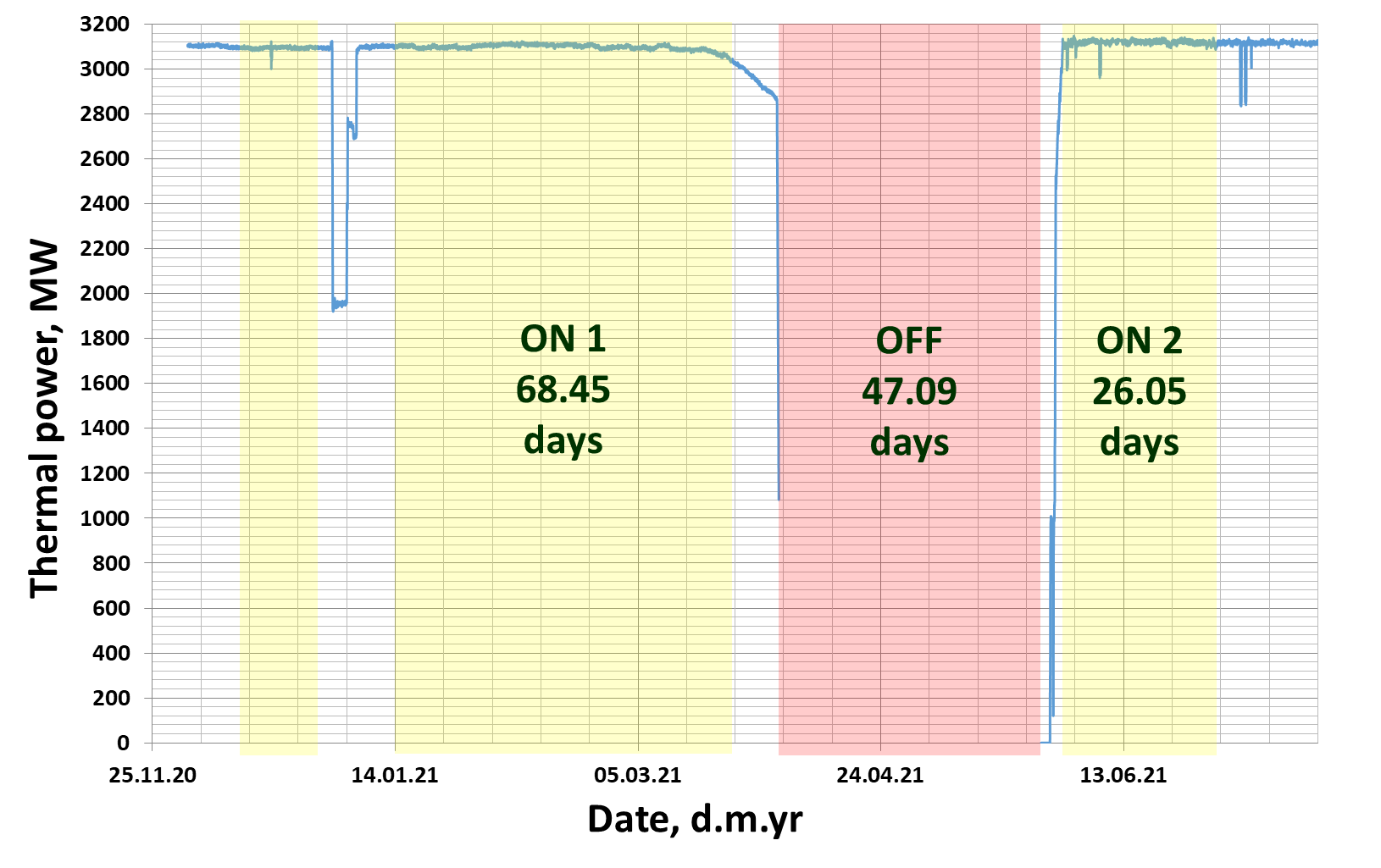}
    \caption{\label{fig:power} Thermal power of the reactor unit 3 during measurements and periods of data taking.}
  \end{center}
\end{figure}

The reactor shut down was for about 60 days in April and May 2021, which allows us to accumulate 47.09 days of statistics with the turned-off reactor. Measurements with reactor ON allowed to accumulate 68.45 days before and 26.05 days after its shut down (see figure~\ref{fig:power}). The periods with temperature changes outside the $\pm$1$^{\circ}$C range and noisy periods were taken out from the analysis.

The reactor antineutrino energy spectra from the reactor has been calculated with help of conversion from electron spectra~\cite{Hub11}, \cite{Haag14}, taking into account the ratio of fusion isotopes~\cite{Kop12}, and the deposed energy per one fission~\cite{Kop04}. A continuous spectrum of up to 10 MeV has been generated. The uncertainties in knowledge of the neutrino spectra are significant for high energy part of the spectra~\cite{Hub11},\cite{Haag14},\cite{Mue11}. Such neutrinos have the biggest impact on expected experimental spectrum from CE{$\mathrm{\nu}$NS. In this work only central values of generated antineutrino spectrum are considered. With the obtained neutrino spectrum and the cross-section (1), the recoil spectrum from coherent neutrino-nucleus scattering has been calculated. The recoil spectra from five stable isotopes of the germanium results in a summation spectrum.

The \nuGeN germanium detector measures only the ionization part of the energy deposition. The ionization part of the energy deposition, the quenching factor, can be described with a Lindhard model~\cite{Lin63} with an adiabatic correction~\cite{Sch16}. There are several predictions and measurements for the quenching factor and for the value of a free parameter $\emph{k}$ of Lindhard model such as~\cite{Sch16},~\cite{Mes95},~\cite{Collar21},~\cite{Bon22}. Depending on the recoil energy the experimental values for the quenching factor are usually in the range of [0.15..0.3]. After applying the quenching factor correction and detector respond the continuous spectrum from \CEvNS with an endpoint of about 0.6 keV has been obtained.

The experimental data with the working reactor together with the generated spectrum from \CEvNS and the ratio of these spectra are shown in figure~\ref{fig:exp2data}.
\begin{figure}
  \begin{center}
    \includegraphics[width=8.6 cm]{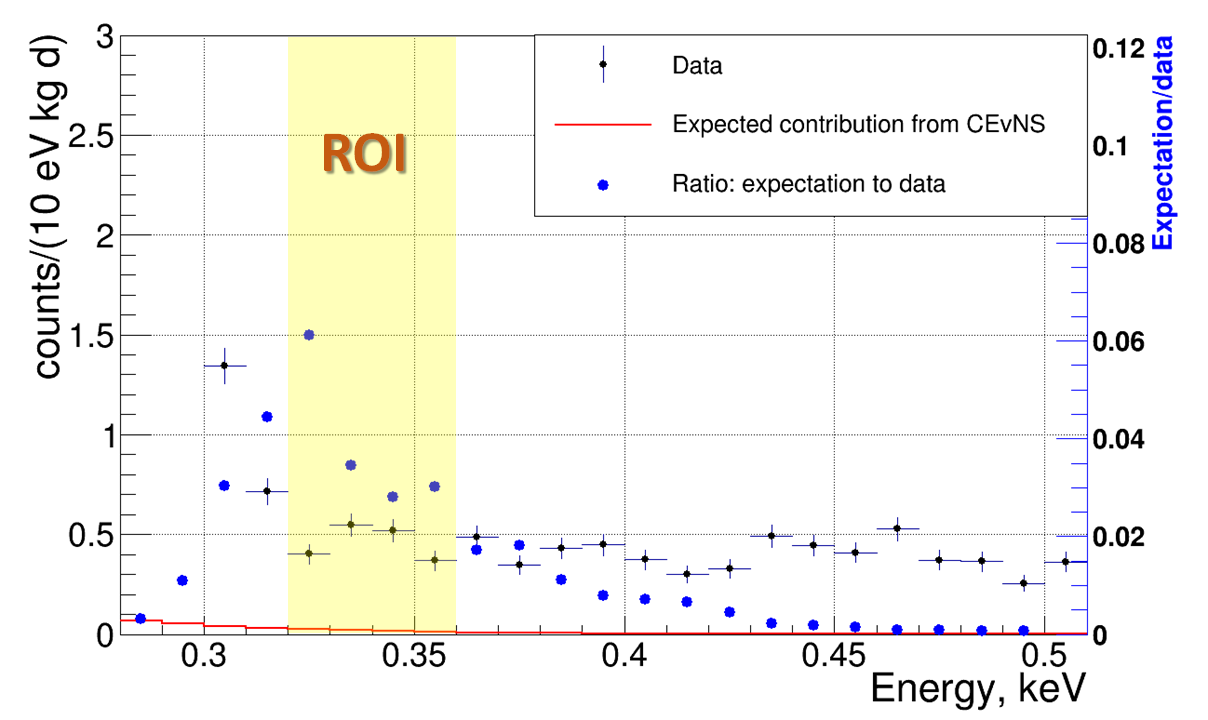}
    \caption{\label{fig:exp2data} Comparison of the spectrum taken with the working reactor with the expected spectrum from CE$\mathrm{\nu}$NS. ROI indicates an energy interval with a maximal sensitivity for CE$\mathrm{\nu}$NS.}
\end{center}
\end{figure}
The expected contribution of the neutrino signal is estimated to be more than 3\% (with the Lindhard model parameter k = 0.179). The Region of Interest (ROI) for the search of \CEvNS of [0.32..0.36] keV has been chosen to have maximum sensitivity to the \CEvNS and avoid the possible influence of noises in the very low energy region.

The comparison of the low energy part of the spectra taken during reactor ON and OFF periods is shown in figure~\ref{fig:on+off}.
\begin{figure}
  \begin{center}
    \includegraphics[width=8.6 cm]{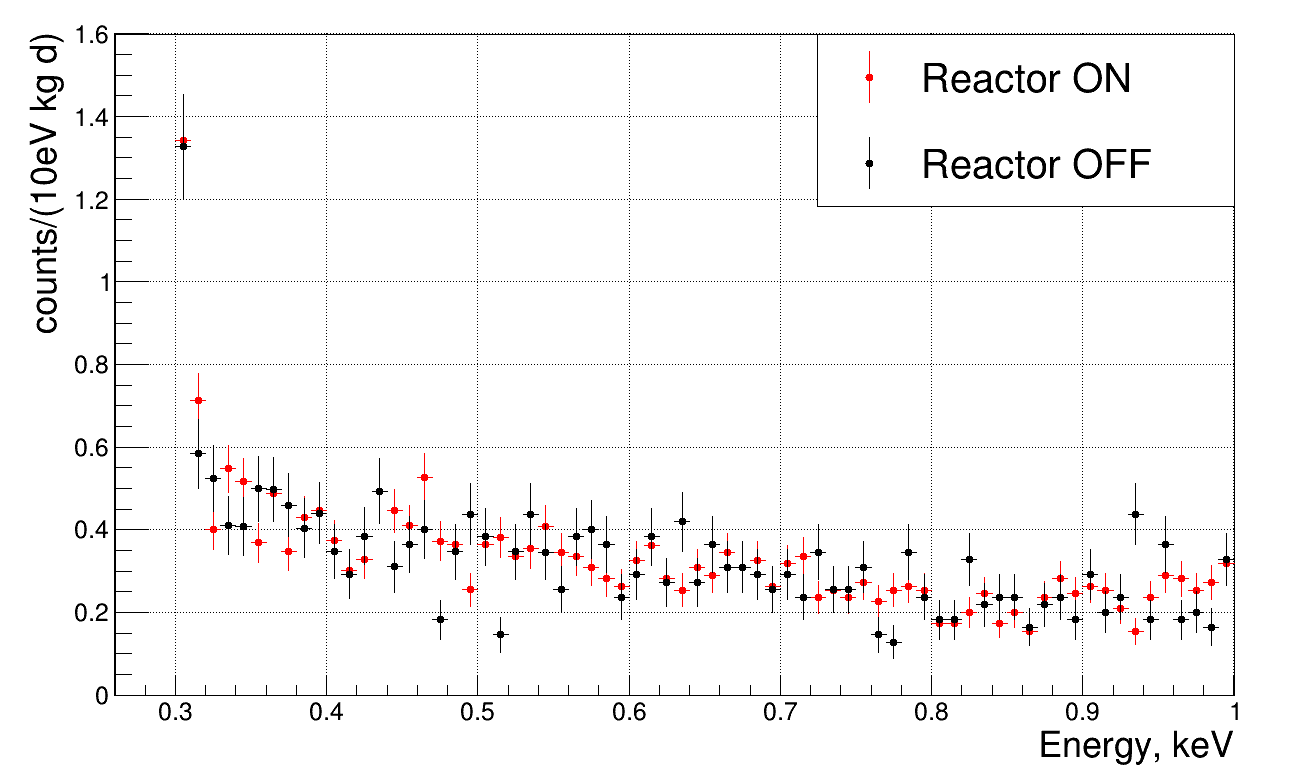}
    \caption{\label{fig:on+off} Low energy part of the spectra taken during reactor ON(red) and reactor OFF(black) periods.}
\end{center}
\end{figure}
No any normalizations besides the measurement time have been applied to the spectra. As one can see from the figure, no significant difference in the background levels and the shape inside and outside of the ROI has been observed during reactor ON and reactor OFF regimes. Therefore, with the current data, there is no hint of the low energy excess expected from CE{$\mathrm{\nu}$NS.

The residual spectrum (ON-OFF) together with the expected spectrum from \CEvNS (with quenching parameter k = 0.179) is shown in figure~\ref{fig:subtract}.
\begin{figure}
  \begin{center}
    \includegraphics[width=8.6 cm]{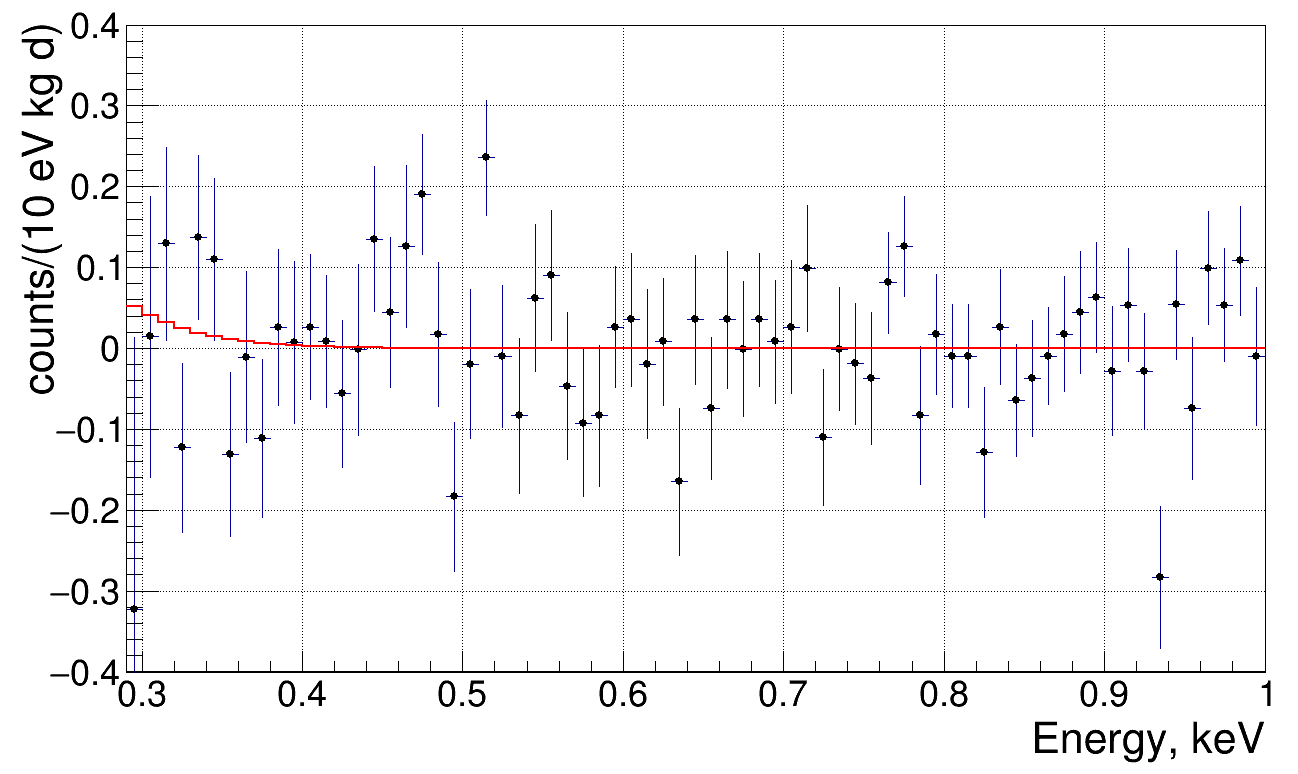}
    \caption{\label{fig:subtract} Residual spectrum (ON-OFF). Red line demonstrates a prediction from \CEvNS with k = 0.179}
  \end{center}
\end{figure}
%
The experimental and expected rates from \CEvNS in the ROI of the spectra are shown in Tab.~\ref{tab:rates}. 
\begin{table}
  \begin{center}
  \caption{\label{tab:rates}Experimental rates in the ROI obtained at \nuGeN and expected contribution from \CEvNS depending on the quenching parameter k.}
\begin{tabular}{lrr}
\toprule
Condition & Measurement  & counts/(kg$\cdot$d) \\
  & time, d & in ROI \\
\hline
Reactor ON   & 94.50 & 2.32$\pm$0.15\\
Reactor OFF  & 47.09 & 2.34$\pm$0.21\\
ON-OFF  & & -0.017$\pm$0.255$\pm$0.03\\
\CEvNS expectation: & & \\
k = 0.3 & & 0.657 \\
k = 0.26 & & 0.415 \\
k = 0.2 & & 0.140 \\
k = 0.179 & & 0.078\\
k = 0.16 & & 0.058\\

\bottomrule
\end{tabular}
  \end{center}
\end{table}
The main systematic uncertainty is expected to be connected with possible shifts in the energy scale and the limited precision of the calibration. Such uncertainty has been estimated by modifying the calibration strategy and changing the parameters of the calibration curve within the uncertainty limits. The estimated systematic uncertainty is expected to be 0.03 counts/(kg$\cdot$d). The obtained difference between rates in the spectra with the Reactor ON and OFF is -0.017$\pm$0.255$\pm$0.03 counts/(kg$\cdot$d). Using a unified approach to the classical statistical analysis~\cite{Feld98} one can put an upper limit for a possible excess in ROI of 0.4 counts/(kg$\cdot$d) with 90\% C.L. The expected contribution from \CEvNS depending on the quenching parameter k is shown in table~\ref{tab:rates}. Comparison of data with the expected rates from \CEvNS disfavor quenching parameters with k $>$ 0.26 with 90\% C.L.  That is similar to result obtained in~\cite{Con21}.

The present work describes the \nuGeN setup. Measurements of 94.50 days with the working reactor and 47.09 days with the turned-off reactor have been performed. The measurements show that achieved background level allows searching for \CEvNS at KNPP. Analysis of the first data shows no significant difference in background level during reactor ON and OFF regimes. No excess at low energy connected with the \CEvNS has been observed. The upper limit on the quenching parameter k $<$ 0.26 with 90\% C.L has been obtained.

The data taking continues, and various effects will be searched within the \nuGeN project. Several new detectors with masses of about 1.0 kg and 1.4 kg will be added to the setup. An internal NaI veto and new electronics will be tested with the aim of background index reduction. A special movable platform has been constructed and commissioned to move the experimental setup toward the reactor core. The distance from the HPGe detector to the center of the reactor core is estimated to be 10.869 m and 11.935 m for higher and lower position, correspondingly. This allows changing the value of the flux by 21\%. All this measures will further improve sensitivity to the search for neutrino scattering.

\begin{acknowledgments}
The authors are grateful to the KNPP directorate and staff for various support and direct help to provide measurements at the reactor site. This work has been partly supported by the Ministry of science and higher education of the Russian Federation (the contract No.075-15-2020-778) and JINR grant for young specialists (No.22-203-02). The work has been supported from European Regional Development Fund-Project "Engineering applications of microworld physics" $\mathrm{(No.~CZ.02.1.01/0.0/0.0/16\_019/0000766)}$.

\end{acknowledgments}

\bibliographystyle{ieeetr}
\bibliography{nuGeN}

\providecommand{\noopsort}[1]{}\providecommand{\singleletter}[1]{#1}%
\begin{thebibliography}{10}

\bibitem{Fre74}
D.~Z. Freedman, ``Coherent effects of a weak neutral current,'' {\em Phys. Rev.
  D}, vol.~9, pp.~1389--1392, Mar 1974.

\bibitem{Dru84}
A.~Drukier and L.~Stodolsky, ``Principles and applications of a neutral-current
  detector for neutrino physics and astronomy,'' {\em Phys. Rev. D}, vol.~30,
  pp.~2295--2309, Dec 1984.

\bibitem{And11}
A.~J. Anderson {\em et~al.}, ``Coherent neutrino scattering in dark matter
  detectors,'' {\em Phys. Rev. D}, vol.~84, p.~013008, Jul 2011.

\bibitem{angleW}
J.~Erler and M.~J. Ramsey-Musolf, ``Weak mixing angle at low energies,'' {\em
  Phys. Rev. D}, vol.~72, p.~073003, Oct 2005.

\bibitem{COHERENT}
D.~Akimov {\em et~al.}, ``Observation of coherent elastic neutrino-nucleus
  scattering,'' {\em Science}, vol.~357, no.~6356, pp.~1123--1126, 2017.

\bibitem{COHERENT2}
D.~Akimov {\em et~al.}, ``First measurement of coherent elastic
  neutrino-nucleus scattering on argon,'' {\em Phys. Rev. Lett.}, vol.~126,
  p.~012002, Jan 2021.

\bibitem{Bed18}
V.~A. Bednyakov and D.~V. Naumov, ``Coherency and incoherency in
  neutrino-nucleus elastic and inelastic scattering,'' {\em Phys. Rev. D},
  vol.~98, p.~053004, Sep 2018.

\bibitem{Rev22}
M.~Abdullah {\em et~al.}, ``Coherent elastic neutrino-nucleus scattering:
  Terrestrial and astrophysical applications, arxiv:2203.07361,'' 2022.

\bibitem{Bed13}
A.~G. Beda {\em et~al.}, ``Gemma experiment: The results of neutrino magnetic
  moment search,'' {\em Physics of Particles and Nuclei Letters}, vol.~10,
  pp.~139--143, Mar 2013.

\bibitem{Bon21}
H.~Bonet {\em et~al.}, ``Large-size sub-kev sensitive germanium detectors for
  the conus experiment,'' {\em The European Physical Journal C}, vol.~81,
  p.~267, Mar 2021.

\bibitem{Bea21}
G.~Beaulieu {\em et~al.}, ``Ricochet progress and status, arxiv:2111.06745,''
  2021.

\bibitem{Ang19}
G.~Angloher {\em et~al.}, ``Exploring with nucleus at the chooz nuclear power
  plant,'' {\em The European Physical Journal C}, vol.~79, p.~1018, Dec 2019.

\bibitem{Ale19}
A.~Aguilar-Arevalo {\em et~al.}, ``Exploring low-energy neutrino physics with
  the coherent neutrino nucleus interaction experiment,'' {\em Phys. Rev. D},
  vol.~100, p.~092005, Nov 2019.

\bibitem{Bel15}
V.~Belov {\em et~al.}, ``{The \ensuremath{\nu}GeN experiment at the Kalinin
  Nuclear Power Plant},'' {\em JINST}, vol.~10, no.~12, p.~P12011, 2015.

\bibitem{Bed07}
A.~G. Beda {\em et~al.}, ``First result for the neutrino magnetic moment from
  measurements with the gemma spectrometer,'' {\em Physics of Atomic Nuclei},
  vol.~70, pp.~1873--1884, Nov 2007.

\bibitem{Col21}
J.~Colaresi {\em et~al.}, ``First results from a search for coherent elastic
  neutrino-nucleus scattering at a reactor site,'' {\em Phys. Rev. D},
  vol.~104, p.~072003, Oct 2021.

\bibitem{Note1}
The detailed investigation of the identical room at the similar reactor core
  unit \#4 are described in~\cite {Ale16}.

\bibitem{Mir20}
``{Mirion Technologies (Canberra Lingolsheim), 1 Chemin de la Roseraie, 67380
  Lingolsheim, France}.''

\bibitem{Cry20}
``{Cryo-Pulse 5 Plus Electrically Refrigerated Cryostat,
  https://www.mirion.com/products/cryo-pulse-5-plus-electrically-refrigerated-cryostat}.''

\bibitem{Vib22}
``{Compact vibration isolation table TS-C30,
  http://tablestable.com/en/products/view/45/}.''

\bibitem{Mor92}
J.~Morales {\em et~al.}, ``{Filtering microphonics in dark matter germanium
  experiments},'' {\em Nucl. Instrum. Meth. A}, vol.~321, pp.~410--414, 1992.

\bibitem{Ste02}
I.~Stekl {\em et~al.}, ``{Present status of the experiment TGV II},'' {\em
  Czech. J. Phys.}, vol.~52, pp.~541--545, 2002.

\bibitem{Hons17}
Z.~Hons and J.~Vl{\'{a}}{\v{s}}ek, ``Data acquisition system for segmented
  reactor antineutrino detector,'' {\em Journal of Instrumentation}, vol.~12,
  pp.~P01022--P01022, jan 2017.

\bibitem{Hons15}
Z.~Hons, ``A versatile daq, monitoring and data processing system for nuclear
  experiments in camac and vme standards, arxiv:1508.01379,'' 2015.

\bibitem{Hub11}
P.~Huber, ``Determination of antineutrino spectra from nuclear reactors,'' {\em
  Phys. Rev. C}, vol.~84, p.~024617, Aug 2011.

\bibitem{Haag14}
N.~Haag {\em et~al.}, ``Experimental determination of the antineutrino spectrum
  of the fission products of $^{238}\mathrm{U}$,'' {\em Phys. Rev. Lett.},
  vol.~112, p.~122501, Mar 2014.

\bibitem{Kop12}
V.~I. Kopeikin, ``Flux and spectrum of reactor antineutrinos,'' {\em Physics of
  Atomic Nuclei}, vol.~75, pp.~143--152, Feb 2012.

\bibitem{Kop04}
V.~I. Kopeikin {\em et~al.}, ``Reactor as a source of antineutrinos: Thermal
  fission energy,'' {\em Physics of Atomic Nuclei}, vol.~67, pp.~1892--1899,
  Oct 2004.

\bibitem{Mue11}
T.~A. Mueller {\em et~al.}, ``Improved predictions of reactor antineutrino
  spectra,'' {\em Phys. Rev. C}, vol.~83, p.~054615, May 2011.

\bibitem{Lin63}
J.~Lindhard {\em et~al.}, ``Integral equations governing radiation effects,''
  {\em Mat. Fys. Medd. Dan. Vid. Selsk}, vol.~33, no.~10, pp.~1--42, 1963.

\bibitem{Sch16}
B.~J. Scholz {\em et~al.}, ``Measurement of the low-energy quenching factor in
  germanium using an $^{88}\mathrm{Y}/\mathrm{Be}$ photoneutron source,'' {\em
  Phys. Rev. D}, vol.~94, p.~122003, Dec 2016.

\bibitem{Mes95}
Y.~Messous and et~al., ``Calibration of a ge crystal with nuclear recoils for
  the development of a dark matter detector,'' {\em Astroparticle Physics},
  vol.~3, no.~4, pp.~361--366, 1995.

\bibitem{Collar21}
J.~I. Collar {\em et~al.}, ``Germanium response to sub-kev nuclear recoils: A
  multipronged experimental characterization,'' {\em Phys. Rev. D}, vol.~103,
  p.~122003, Jun 2021.

\bibitem{Bon22}
A.~Bonhomme {\em et~al.}, ``Direct measurement of the ionization quenching
  factor of nuclear recoils in germanium in the kev energy range,
  arxiv:2202.03754,'' 2022.

\bibitem{Feld98}
G.~J. Feldman and R.~D. Cousins, ``Unified approach to the classical
  statistical analysis of small signals,'' {\em Phys. Rev. D}, vol.~57,
  pp.~3873--3889, Apr 1998.

\bibitem{Con21}
H.~Bonet {\em et~al.}, ``Constraints on elastic neutrino nucleus scattering in
  the fully coherent regime from the conus experiment,'' {\em Phys. Rev.
  Lett.}, vol.~126, p.~041804, Jan 2021.

\bibitem{Ale16}
I.~Alekseev {\em et~al.}, ``{DANSS: Detector of the reactor AntiNeutrino based
  on Solid Scintillator},'' {\em JINST}, vol.~11, no.~11, p.~P11011, 2016.

\end{thebibliography}
\end{document}